\documentclass[aps,pra,twocolumn,showpacs,preprintnumbers,amsmath,amssymb,superscriptaddress,10pt]{revtex4-1}

\usepackage{amsmath}  
\usepackage{amsfonts} 
\usepackage{graphicx} 
\usepackage{amssymb}
\usepackage[utf8]{inputenc}
\usepackage{dcolumn}
\usepackage{bm}
\usepackage{graphicx}
\usepackage{tikz}
\usepackage[caption=false]{subfig}

\newcommand{\diff}{\text{d}}

\renewcommand{\Re}{\mathrm{Re}}
\renewcommand{\Im}{\mathrm{Im}}

\newcommand{\vek}[1]{\boldsymbol{\mathbf{#1}}}
\newcommand{\vekh}[1]{\hat{{\boldsymbol{\mathbf{#1}}}}}

\newcommand{\be}{\begin{equation}}
\newcommand{\ee}{\end{equation}}
\newcommand{\mc}[1]{\vek{\mathcal{#1}}}

\newcommand{\mull}{\mu_{\text{ll}}}
\newcommand{\mutl}{\mu_{\text{tl}}}
\newcommand{\mumm}{\mu_\text{mm}}
\newcommand{\muvy}{\mu_\text{vy}}

\newcommand{\Mvy}{\mc{M}^\text{vy}}

\newcommand{\epsOK}{\epsilon(\omega,\vek{k})}
\newcommand{\epsOKv}{\vek\epsilon(\omega,\vek{k})}

\newcommand{\epseff}{\epsilon}
\newcommand{\mueff}{\mu}
\newcommand{\e}[1]{\text{exp}({#1})}
\newcommand{\kB}{\vek{k}_\text{B}}
\newcommand{\Mmm}{\mc{M}^\text{mm}}

\newcommand{\rTM}{r_{\text{TM}}}
\newcommand{\bigO}{\mathcal O}
\renewcommand{\e}[1]{\text{e}^{#1}}

\begin{document}

	\title{The magnetic permeability in Fresnel's equation}
	
	\author{Hans Olaf H\aa genvik}\email{hansolaf@me.com}
	\affiliation{Department of Electronic Systems, NTNU -- Norwegian University of Science and Technology, NO-7491 Trondheim, Norway}
	
	\author{Johannes Skaar}\email{johannes.skaar@its.uio.no}
	\affiliation{Department of Technology Systems, University of Oslo, P.O. Box 70, NO-2027 Kjeller, Norway}
	




\begin{abstract}
Magnetic permeabilities derived for infinite, periodic media are used in the Fresnel equation to calculate the reflection from corresponding  semi-infinite media. By comparison to finite-difference-time-domain (FDTD) simulations, we find that the Fresnel equation gives accurate results for 2D metamaterials which mimic natural magnetism, in a frequency range where the magnetic moment density dominates the $\bigO(k^2)$ part of the total Landau--Lifshitz permittivity. For a 1D layered structure, or for large frequencies, the correspondence is poor. We also demonstrate that even if a medium is described accurately by a local permittivity and permeability, the Fresnel equation is not necessarily valid. 
\end{abstract}
	
\maketitle


\section{Introduction}
When an electromagnetic wave is incident from vacuum to a plane boundary, the Fresnel equations can often be used to calculate the reflection and transmission coefficients, provided the medium can be described by a permittivity $\epsilon$ and permeability $\mu$. Clever design of composite media with structures much smaller than the wavelength, so called metamaterials, has made it possible to tailor the effective parameters $\epsilon$ and $\mu$ to obtain novel reflection and transmission properties. A nontrivial permeability $\mu$ may for instance be obtained through metallic inclusions, where circulating currents may mimic the magnetic moment in actual atoms. Recent progress in nanostructuring techniques makes this possible even at optical frequencies, where the magnetic response for natural media is absent \cite{pendry99,Shelby01,pendry03b,pendry04,enkrich05,alu06}.

For most natural media the electric and magnetic dipole approximations provide a sufficient description. For metamaterials, however, it turns out that the electric quadrupole moment \cite{zhang2008}, or even higher order multipoles \cite{Dirdal18} may be significant. It can be shown that magnetism is a second order spatially dispersive effect of the Landau--Lifshitz total permittivity \cite{Landau-Lifshits04}. This is also the case for the electric quadrupole density, and even parts of the electric octupole and magnetic quadrupole density \cite{Dirdal18}. If we include induced magnetic moment in our medium model, these other terms should also be taken into account. 

The effective constitutive parameters of an infinite, periodic metamaterial are retrieved by inserting a source in the medium, and calculating the resulting, microscopic fields and currents \cite{silveirinha07,alu11,Yaghjian2013374,skaar18}. This method is sometimes called \textit{current driven homogenization}. It is important to note that the parameters are obtained for an infinite structure, and it is not obvious how well they describe the medium response when a boundary is present. 

The derivation of Fresnel's equations \cite{Landau-Lifshits04} requires the medium to be characterized \emph{only} by a set of local permittivity and local permeability. In other words, these parameters should not only be independent of the wavevector $\vek k$, but they should also accurately model the entire electromagnetic response of the medium. This means that additional terms (such as higher order multipoles) should not be necessary to represent the induced current in the unit cells. The derivation of Fresnel's equations also requires the tangential components of the macroscopic electric and magnetic fields to be continuous. Whether or not these fields are continuous will depend on the particular definition of macroscopic fields in the homogenization method. None of the conditions are necessarily satisfied for a homogenized metamaterial.

Although there exist boundary conditions for certain weakly spatially dispersive media \cite{Golubkov1995, raab13,yaghjian14,Silveirinha14,yaghjian16}, it is generally unclear how the higher order multipoles affect the reflection coefficient of a semi-infinite metamaterial. Still it is possible to calculate the reflection from a semi-infinite metamaterial using only results calculated for an infinite structure \cite{joannopoulos}. This is done by matching the Bloch modes of the periodic metamaterial to the incident and reflected plane waves (Sec. \ref{sec:semiinfinit}). The Bloch modes are valid arbitrarily close to the boundary. Thus, in this method one may argue that there is no ``transition layer'' with different constitutive behavior \footnote{For 1D there is only a single, propagating Bloch mode in the metamaterial. This mode is present for all $z>0$; thus there is no ``transition layer''. For 2D or 3D, there will also be modes with complex propagation constants, decaying exponentially away from the boundary. One view is that the region where the complex modes are significant, is a ``transition layer''. Another view is that all modes, no matter if the propagation constants are real or complex, are present in the entire half-space; thus there is no ``transition layer''.}.

In this work we insert permittivities and permeabilities obtained from infinite structures in the Fresnel equation, and see how well they predict the reflection from a semi-infinite metamaterial. We compare the results with the ``exact'' solution obtained from finite-difference-time-domain (FDTD) simulations of the same structure. We do this for three different periodic media: a 1D layered dielectric structure, a 2D metal bar structure and a 2D metallic split-ring medium. For  comparison, we use four different definitions of magnetic permeability \cite{skaar18}; the conventional definition using the magnetic moment density \cite{Landau-Lifshits04,alu11} $\mumm$, a definition based on a similar integral \cite{vinogradov1999,Yaghjian2013374} $\muvy$, a definition based on the $\bigO(k^2)$ term of the Landau--Lifshitz permittivity \cite{Landau-Lifshits04,silveirinha07,skaar18} $\mutl$, and a trivial permeability $\mull=1$.

A similar analysis limited to a 1D layered structure was done in Ref. \cite{Markel13}. An expression for the reflection coefficient was derived using the transfer matrix method. This exact solution was compared with the reflection coefficient calculated using $\mutl$ from current driven homogenization. The reflection coefficient was predicted accurately only in the zero-frequency limit, where $\mutl=1$. In the frequency range where the homogenization method predicted a nontrivial magnetic response, the reflection coefficient was not more accurate with current driven homogenization than by simply setting $\mutl=1$. This suggests that the permeability obtained from current driven homogenization is useless in predicting reflection from a 1D layered structure. 

These 1D results are confirmed here. However, by considering 2D metamaterials that mimic natural magnetism, we find that the permeabilities from current driven homogenization predict the reflection fairly well, in a frequency range where the permeabilities are nontrivial. From the numerical results it is suggested that this frequency range is where the three nontrivial permeabilities coincide, i.e., where the magnetic moment density dominates the $\bigO(k^2)$ part of the Landau--Lifshitz total permittivity (Sec. \ref{sec:examples}).

For higher frequencies, the reflection is not accurately predicted by the parameters from current driven homogenization. Clearly, the reason can be nonlocal $\epsilon$ and $\mu$, and/or that they do not alone represent the total induced current in the unit cells. However, we demonstrate that even if the medium is described accurately by local $\epsilon$ and $\mu$, the reflection may not be predicted well by Fresnel's equation. This is due to the macroscopic fields being defined as fundamental Floquet modes on each side of the boundary, making the tangential electric and magnetic fields discontinuous. The macroscopic fields could have been defined differently, e.g. using test function averaging \cite{russakoff,jackson,silveirinha09}. This would make the fields continuous, at the expense of introducing a ``transition layer'' close to the boundary, with different constitutive relations.

\section{Infinite periodic medium}\label{sec:four_mus}
Current driven homogenization \cite{silveirinha07,alu11,Yaghjian2013374,skaar18} considers an \textit{infinite} periodic structure, which is excited by a source. It is convenient to look at a single spatial Fourier component of the source, $\mc J_\text{ext}(\vek{r}) = \vek{J}_0 \e{i\vek{k}\cdot \vek{r}}$, where $\vek{J}_0$ is independent of $\vek{r}$, and $\vek{k}$ is a fixed parameter, in general unrelated to frequency $\omega$. The microscopic electric field $\vek{e}(\vek r)$ is found from the microscopic Maxwell equations. In our case we use the finite-difference-frequency-domain (FDFD) method  \cite{costa09,Dirdal18}. The microscopic, induced current density is given by 
\begin{equation}
    \vek{j}(\vek r) = -i\omega\epsilon_0[\varepsilon(\vek{r})-1]\vek{e}(\vek{r}),
\end{equation}
where $\varepsilon(\vek{r})$ is the microscopic (relative) permittivity inside the unit cell, and $\epsilon_0$ is the vacuum permittivity.

The spatial dependency $\e{i\vek{k}\cdot\vek{r}}$ of the source will cause the microscopic field to be Bloch-periodic with Bloch-vector equal to $\vek k$:
\begin{equation}\label{emic}
    \vek{e}(\vek{r}) = \vek{u}_{\vek e}(\vek{r}) \e{i\vek{k}\cdot\vek{r}}.
\end{equation}
Here $ \vek{u}_{\vek e}(\vek{r})$ is a periodic function with the same periodicity as the lattice. The other microscopic fields are expressed similarly. Let $V$ be the unit cell containing the origin. The macroscopic, electric field is defined as the fundamental Floquet mode, expressed from the averaged Bloch function \cite{silveirinha07}:
\begin{equation}\label{Emac}
    \mc{E}(\vek{r}) = \frac{\e{i\vek{k}\cdot\vek{r}}}{V}\int_V \vek{e}(\vek{r}) \e{-i\vek{k}\cdot \vek{r}}\diff^3r.
\end{equation}
The macroscopic magnetic field $\mc B(\vek r)$ and the macroscopic induced current $\mc J(\vek r)$ are defined similarly. For simplicity of notation, we will suppress the $\vek r$ dependence of these fields. Homogenizing the microscopic Maxwell equations using \eqref{Emac}, we obtain \cite{silveirinha07,alu11,Yaghjian2013374,skaar18}
\begin{subequations}\label{metaME}
\begin{align}
    i\vek{k}\times \mc{E} &= i\omega \mc{B},\\
    i\vek{k}\times \mc{H} &= -i\omega\mc{D} + \mc{J}_\text{ext},
\end{align}
\end{subequations}
where 
\begin{subequations}\label{DHgen}
\begin{align}
\mc D &= \epsilon_0\mc E + \frac{\mc J- i\vek k\times\mc M}{-i\omega}, \label{Ddefgen}\\
\mc H &= \mc B/\mu_0-\mc M,
\end{align}
\end{subequations}
and $\mu_0$ is the vacuum permeability. Here, $\mc M$ is in principle arbitrary, as evident by substituting \eqref{DHgen} into \eqref{metaME}. 

Below we will consider four possible definitions of $\mc M$, leading to four associated permeabilities \cite{skaar18}. Let the medium be 2D, as in the $z>0$ region of Fig. \ref{fig:setup}. Assume TM polarization with microscopic magnetic field in the $\vekh y$-direction, and source and microscopic electric field perpendicular to $\vekh y$. We consider media with a center of symmetry, and let the origin be this center of symmetry: $\varepsilon(\vek r)=\varepsilon(-\vek r)$. This is done to ensure gyrotropic effects are negligible, as we are interested in how the magnetic response contribute to reflections from metamaterial boundaries.
\begin{enumerate}
    \item Landau--Lifshitz (ll) formulation \cite{Landau-Lifshits04}: 
    \begin{equation}
        \mc M^{\text{ll}} = 0.
    \end{equation}
    This means that the permeability is trivial:
    \begin{equation}
        \vek\mu_\text{ll} = 1, \label{trivmu}
    \end{equation}
    and 
    \begin{equation}\label{epsLLDJ}
        \mc{D}^\text{ll} = \epsilon_0\mc{E} + \frac{\mc{J}}{-i\omega} = \epsilon_0\vek\epsilon(\omega,\vek{k})\mc{E}.
    \end{equation}
    Here $\vek\epsilon(\omega,\vek k)$ is the Landau--Lifshitz (relative) permittivity tensor, describing the entire electromagnetic response of the infinite, periodic medium. We can describe $\vek\epsilon(\omega,\vek k)$ to second order in $k$:
    \begin{equation}\label{epsOKk2}
        \epsilon_{ij}(\omega,\vek k) = \epsilon_{ij} + \beta_{iklj}k_kk_l,
    \end{equation}
    where summation over repeating indices is implied. In general the tensor coefficients $\beta_{iklj}$ are allowed to be dependent on $\vek k$, to describe the remainder of the Taylor expansion. However, for media and frequency ranges with weak spatial dispersion, the higher order terms are negligible, and $\beta_{iklj}$ is constant.
    
    \item Magnetic moment (mm) formulation \cite{Landau-Lifshits04,alu11}: 
    \begin{equation} 
        \Mmm =     \frac{\e{i\vek{k}\cdot\vek{r}}}{2V}\int_V \vek{r}\times\vek{j}(\vek{r}) \diff^3r.
    \end{equation}
    Assuming $\vek k=k\vekh z$, and TM, we can define a magnetic permeability $\mu_\text{mm}$ in the $\vekh y$-direction by \cite{skaar18} 
    \begin{equation} \label{mumm}
        \mathcal M^\text{mm} = \frac{1}{\mu_0\omega} \nu_{yzz}k\mathcal E_z + \frac{1}{\mu_0}(1-\mu_\text{mm}^{-1})\mathcal B.
    \end{equation}
    Here $\nu_{yzz}$ is a constant (constitutive parameter).
    
    \item Vinogradov--Yaghjian (vy) formulation \cite{vinogradov1999,Yaghjian2013374}:
    \begin{equation}
        \Mvy = \frac{\e{i\vek{k}\cdot\vek{r}}}{2V}\int_V \vek{r}\times\vek{j}(\vek{r}) \e{-i\vek{k}\cdot\vek{r}} \diff^3r.
    \end{equation}
    The permeability $\mu_\text{vy}$ is defined from $\Mvy$ in the same way as $\mu_\text{mm}$ is defined from $\Mmm$.
    
    \item Transversal--longitudinal formulation \cite{Landau-Lifshits04,silveirinha07,skaar18}: Here the permeability is defined from the second order $\bigO(k^2)$ part of the Landau--Lifshitz permittivity \eqref{epsOKk2},
    \begin{equation}\label{mutlgen}
        \left[1-\vek\mu_{\text{tl}}^{-1}\right]_{mn} = \varepsilon_{mip}\varepsilon_{njq}\frac{k_kk_lk_pk_q}{k^4}\beta_{iklj}.
    \end{equation}
    We are interested in the permeability for the $y$-direction, with $\vek k=k\vekh z$:
    \begin{equation}\label{muCDH}
        1-\mu_{\text{tl}}^{-1} = \beta_{xzzx}.
    \end{equation}
\end{enumerate}
For all cases, a permittivity is defined by
\begin{equation}\label{epsCDH}
    \vek\epsilon = \lim_{\vek{k}\to 0} \vek\epsilon(\omega,\vek k).
\end{equation}
We will mainly be interested in the permittivity for the $x$-direction $(\vek\epsilon)_{xx}$, which will be denoted $\epsilon$.

In the special case where a medium is described accurately by (only) a permittivity $\vek\epsilon$ and some permeability $\vek\mu$, we have
\begin{subequations}\label{DHspec}
\begin{align}
\mc D &= \epsilon_0\vek{\epsilon} \mc E, \\
\mc H &= \mu_0^{-1}\vek\mu^{-1} \mc B.
\end{align}
\end{subequations}
Comparing to \eqref{DHgen} and eliminating $\mc M$, we find $\mc J$, which can be substituted into \eqref{epsLLDJ} to find
\begin{equation}\label{epsOK}
    \epsOKv = \vek\epsilon - \frac{c^2}{\omega^2}\vek{k}\times \big[ 1 - \vek\mu^{-1}\big]\times\vek{k}.
\end{equation}
This is the Landau--Lifshitz permittivity corresponding to $\vek\epsilon$ and $\vek\mu$. We note that the permeability $\vek\mu$ can be seen as a $\bigO(k^2)$ effect of $\epsOKv$. If $\vek\epsilon$ and $\vek\mu$ are independent of $\vek k$, and \eqref{epsOK} is valid, the medium will be called \emph{local}.

We have defined four permeabilities $\mull=1$, $\mumm$, $\muvy$, and $\mutl$, and one permittivity \eqref{epsCDH}. The connection between the permeabilities, and properties in general, are discussed in detail in Ref. \cite{skaar18}, including causality, passivity, symmetry, asymptotic behavior, and origin dependence. The analysis in Ref. \cite{skaar18} covers an infinite periodic structure. We are now interested in the semi-infinite case. Will the constitutive parameters above be useful to calculate the reflection coefficient using the Fresnel equation?

\section{Semi-infinite medium}\label{sec:semiinfinit}
Consider a boundary surface between vaccuum and a 2D periodic metamaterial. The metamaterial consists of square unit cells with lattice parameter $a$, and covers the semi-infinite region $z>0$, extending infinitely in the $x$-direction. This medium is illuminated by a normally incident TM-polarized plane wave, using a source located somewhere to the left of the boundary. We are interested in the reflected fields. For simplicity we still assume that the unit cells consist of non-magnetic constituents, and that the metamaterial has a center of symmetry when viewed as an infinite periodic medium.

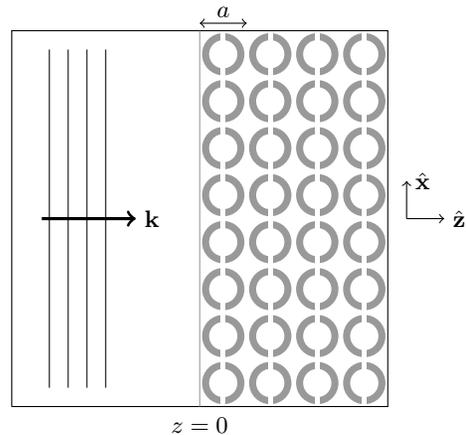
\begin{figure}[!htb]
    \center
    \begin{tikzpicture} [scale=5]
        \foreach \x in {0.0625,0.1875,0.3125,0.4375} 
             \foreach \y in {-0.0625,-0.1875,-0.3125,-0.4375,0.0625,0.1875,0.3125,0.4375}  
            { 
                \path [draw=none,fill=black!40!white, fill opacity = 1] (\x,\y) circle (0.0563);
                \path [draw=none,fill=white, fill opacity = 1] (\x,\y) circle (0.0375);
            } 
        \foreach \x in {0.0625,0.1875,0.3125,0.4375} 
            {
                \fill[black!0!white] (\x-0.00625,-0.5) rectangle (\x+0.00625,0.5);
            }
        \draw[-] (-0.5,-0.5)--(-0.5,0.5)--(0.5,0.5)--(0.5,-0.5)--(-0.5,-0.5);
        \draw[-,black!40!white] (0,-0.5)--(0,0.5);
        \node [] at (0,-0.55) {$z=0$};
        \foreach \x in {-0.4, -0.35, -0.3, -0.25} 
        {
           \draw[-] (\x,-0.45)--(\x,0.45);
        }
        \draw[<->] (0,0.52)--(0.125,0.52);
        \node [] at (0.0625,0.55) {$a$};
        \draw[->,very thick] (-0.42,0)--(-0.17,0);
        \node [right] at (-0.17,0) {$\vek k$};
        \draw[->] (0.55,0)--(0.65,0);
        \node [right] at (0.65,0) {$\vekh z$};
        \draw[->] (0.55,0)--(0.55,0.1);
        \node [right] at (0.55,0.1) {$\vekh x$};
    \end{tikzpicture}
    \caption{A semi-infinite 2D periodic metamaterial covers the region $z>0$. For $z<0$ there is vacuum. The structure is illuminated by a TM plane wave source located somewhere to the left of the boundary. The square unit cells of the metamaterial have lattice parameter $a$.}
    \label{fig:setup}
\end{figure}

Considering a conventional, homogeneous medium rather than the metamaterial, the reflection coefficient would usually be given by the Fresnel equation for a TM-polarized wave:
\begin{equation}
        \rTM = \frac{\epsilon k_0 - k}{\epsilon k_0 + k}. \label{rTM}\\
\end{equation}
Here $\rTM$ is defined as the ratio between the complex amplitudes of reflected and incident magnetic $\mathcal B$-field at the boundary. Moreover $k_{0}=\omega/c$ and $k=\sqrt{\epsilon\mu}\omega/c$ are the wavenumbers in vacuum and in the medium, respectively, where $\epsilon$ and $\mu$ are the electromagnetic parameters of the medium. If the medium is anisotropic, \eqref{rTM} still applies if the permittivity and permeability tensors are diagonal in the coordinate system shown in Fig. \ref{fig:setup}. Then, in \eqref{rTM}, $\epsilon$ refers to the permittivity in the $x$-direction, and $\mu$ to the permeability in the $y$-direction.

The Fresnel equation is derived from Maxwell's equations for local media. For \eqref{rTM} to predict the reflection of the semi-infinite metamaterial accurately, we should therefore require $\vek\epsilon$ and $\vek\mu$ to be independent of $\vek{k}$, \emph{and} \eqref{epsOK} to be valid. A necessary condition is that $\vek\epsilon(\omega,\vek k)$ is described accurately by a second order Taylor expansion \eqref{epsOKk2}, with $\beta_{iklj}$ independent of $\vek k$. We should also require certain elements of $\beta_{iklj}$ to vanish (or be small), such that \eqref{epsOKk2} can be written in the form \eqref{epsOK}. The permeability $\vek\mu_\text{tl}$ will then describe the $\bigO(k^2)$ part of $\vek\epsilon(\omega,\vek k)$ accurately \cite{Landau-Lifshits04,silveirinha07,skaar18}.

For a semi-infinite periodic medium, such as that in Fig. \ref{fig:setup}, we can calculate the reflection coefficient using a mode matching technique (Appendix \ref{app:newRefl}), or with FDTD (Sec. \ref{sec:examples}). The mode matching technique is particularly simple in 1D when the metamaterial is homogeneous in the $x$ and $y$ directions. Assuming there is a vacuum layer immediately to the right of the boundary, as in Fig. \ref{fig:setup} and our numerical examples, the reflection coefficient becomes (Appendix \ref{app:newRefl}):
\begin{equation}\label{newRefl_copy}
 \rTM = \frac{k_0-k_\text{B}+i\frac{u_b'(0)}{u_b(0)}}{k_0+k_\text{B}-i\frac{u_b'(0)}{u_b(0)}}.
\end{equation}
Here $u_b(z)$ is the periodic Bloch function for the magnetic $b(z)$ field. For a layered structure the Bloch wavenumber $k_\text{B}$ is given by the dispersion relation \cite{Markel13}
\begin{align}\label{Blochk}
    \cos(k_\text{B}a) = &\cos\big[(n_1d_1 + n_2d_2)\omega a/c\big] \\&- \frac{(n_1-n_2)^2}{2n_1n_2}\sin\big( n_1d_1\omega a/c\big)\sin\big( n_2d_2\omega a/c\big)\nonumber,
\end{align}
where $d_1$ and $d_2$ are the layer thicknesses ($d_1+d_2=a$), as seen in the upper part of Fig. \ref{fig:UnitCells}. For our examples, $n_1=1$ and $n_2=\sqrt{\epsilon}$.

\begin{figure}[!t]
\center
\includegraphics[width=0.4\textwidth]{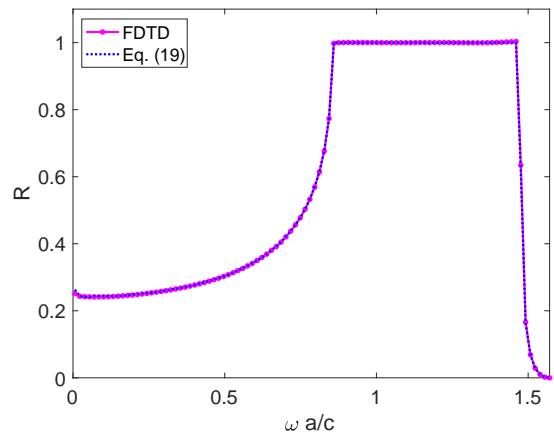}
\caption{Power reflection $R=|\rTM|^2$ obtained from FDTD, and from \eqref{newRefl_copy}, where $k_\text{B}$ is obtained from \eqref{Blochk}, and $u_b(z)$ is obtained from $u_b(z) = b(z)\e{-ik_\text{B}z}$. The microscopic field $b(z)$ can be obtained e.g. using a transfer matrix method, or (as here) with FDFD. The unit cell was represented by a grid of $200\times 1$ and $5000 \times 1$ points in the FDTD and FDFD calculations, respectively.}
\label{fig:layers_new_refl}
\end{figure}

The power reflection coefficient $R=|\rTM|^2$, with $\rTM$ given by \eqref{newRefl_copy}, is plotted in Fig.~\ref{fig:layers_new_refl} for the upper metamaterial unit cell in Fig.~\ref{fig:UnitCells}. Note that the reflection coefficient is calculated solely by quantities describing propagation in an infinite periodic structure. As seen in the figure, the reflection coefficient matches that from an independent FDTD calculation.

The mode matching technique in 1D leads to a very interesting result (Appendix \ref{app:newRefl}): If we use \eqref{Emac} to define the macroscopic fields for the incident, reflected and transmitted wave separately (using the corresponding wavevectors), the macroscopic fields are well defined for all $z\neq 0$. For example, considering $z>0$, the $z$-dependence becomes $\exp(ik_\text{B}z)$ everywhere. In other words, there is no ``transition layer'' in the vicinity of the interface with different behavior. This comes at a price: The tangential macroscopic fields are not continuous across the boundary. Thus the Maxwell boundary conditions do not apply [see \eqref{disE}-\eqref{disB}], unless we define (somewhat artificial) macroscopic electric and magnetic surface currents.

In 2D a simple expression such as \eqref{newRefl_copy} cannot be found for the reflection coefficient. This is because additional Bloch modes will be excited at the boundary. For small frequencies these modes have usually complex propagation constants, meaning that they are present close to the boundary. To find the reflection coefficient, the full mode matching matrix problem must be solved. Alternatively, one can use a numerical method such as FDTD.

\section{Comparison of reflection coefficients from FDTD and Fresnel's equation}\label{sec:examples}
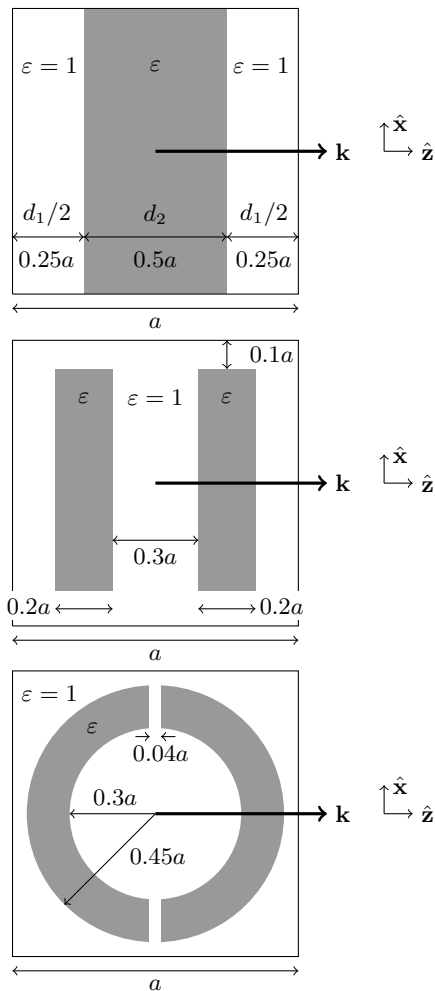
\begin{figure}[!t]
\center
\begin{tikzpicture} [scale=3.8]
\fill[black!40!white] (-0.25,-0.5) rectangle (0.25,0.5);
\draw[-] (-0.5,-0.5)--(-0.5,0.5)--(0.5,0.5)--(0.5,-0.5)--(-0.5,-0.5);
\node [] at (0,0.3) {$\varepsilon$};
\node [] at (-0.37,0.3) {$\varepsilon=1$};
\node [] at (0.37,0.3) {$\varepsilon=1$};
\draw[<->] (-0.25,-0.3)--(0.25,-0.3);
\node at (0,-0.38) {$0.5a$};
\node at (0,-0.22) {$d_2$};
\draw[<->] (-0.5,-0.3)--(-0.25,-0.3);
\node at (-0.38,-0.38) {$0.25a$};
\node at (-0.38,-0.22) {$d_1/2$};
\draw[<->] (0.25,-0.3)--(0.5,-0.3);
\node at (0.38,-0.38) {$0.25a$};
\node at (0.38,-0.22) {$d_1/2$};
\draw[->,very thick] (0,0)--(0.6,0);
\node [right] at (0.6,0) {$\vek k$};
\draw[<->] (-0.5,-0.55)--(0.5,-0.55);
\node [] at (0,-0.6) {$a$};
\draw[->] (0.8,0)--(0.9,0);
\node [right] at (0.9,0) {$\vekh z$};
\draw[->] (0.8,0)--(0.8,0.1);
\node [right] at (0.8,0.1) {$\vekh x$};
\fill[black!0!white] (-0.79,0) rectangle (-0.8,0.1);
\end{tikzpicture}

\begin{tikzpicture} [scale=3.8]
\fill[black!40!white] (0.15,-0.4) rectangle (0.35,0.4);
\fill[black!40!white] (-0.35,-0.4) rectangle (-0.15,0.4);
\draw[-] (-0.5,-0.5)--(-0.5,0.5)--(0.5,0.5)--(0.5,-0.5)--(-0.5,-0.5);
\fill[black!0!white] (-0.52,-0.48) rectangle (0.52,-0.38);
\node [] at (0,0.3) {$\varepsilon=1$};
\node [] at (-0.25,0.3) {$\varepsilon$};
\node [] at (0.25,0.3) {$\varepsilon$};
\draw[<->] (-0.15,-0.2)--(0.15,-0.2);
\node [below] at (0,-0.2) {$0.3a$};
\draw[<->] (-0.35,-0.44)--(-0.15,-0.44);
\node [left] at (-0.335,-0.43) {$0.2a$};
\draw[<->] (0.35,-0.44)--(0.15,-0.44);
\node [right] at (0.335,-0.43) {$0.2a$};
\draw[->,very thick] (0,0)--(0.6,0);
\node [right] at (0.6,0) {$\vek k$};
\draw[<->] (-0.5,-0.55)--(0.5,-0.55);
\node [] at (0,-0.6) {$a$};
\draw[<->] (0.25,0.4)--(0.25,0.5);
\node [right] at (0.3,0.45) {$0.1 a$};
\draw[->] (0.8,0)--(0.9,0);
\node [right] at (0.9,0) {$\vekh z$};
\draw[->] (0.8,0)--(0.8,0.1);
\node [right] at (0.8,0.1) {$\vekh x$};
\fill[black!0!white] (-0.79,0) rectangle (-0.8,0.1);
\end{tikzpicture}

\begin{tikzpicture} [scale=3.8]
\path [draw=none,fill=black!40!white, fill opacity = 1] (0,0) circle (0.45);
\path [draw=none,fill=white, fill opacity = 1] (0,0) circle (0.30);
\draw[-] (-0.5,-0.5)--(-0.5,0.5)--(0.5,0.5)--(0.5,-0.5)--(-0.5,-0.5);
\node [] at (-0.22,0.30) {$\varepsilon$};
\node [] at (-0.37,0.42) {$\varepsilon=1$};
\draw[<->] (-0.5,-0.55)--(0.5,-0.55);
\node [] at (0,-0.6) {$a$};
\draw[->] (0,0)--(-0.30,0);
\node [above] at (-0.14,0) {$0.3a$};
\draw[->] (0,0)--(-0.3182,-0.3182);
\node [right] at (-0.12,-0.15) {$0.45a$};
\draw[->,very thick] (0,0)--(0.6,0);
\node [right] at (0.6,0) {$\vek k$};
\fill[black!0!white] (-0.02,-0.48) rectangle (0.02,-0.29);
\fill[black!0!white] (-0.02,0.29) rectangle (0.02,0.48);
\draw[->] (-0.07,0.27)--(-0.02,0.27);
\draw[->] (0.07,0.27)--(0.02,0.27);
\node [] at (0.02,0.21) {$0.04a$};
\draw[->] (0.8,0)--(0.9,0);
\node [right] at (0.9,0) {$\vekh z$};
\draw[->] (0.8,0)--(0.8,0.1);
\node [right] at (0.8,0.1) {$\vekh x$};
\fill[black!0!white] (-0.79,0) rectangle (-0.8,0.1);
\end{tikzpicture}

\caption{Different unit cells for the simulations: Layered structure (upper); two bars (middle); split-ring resonator (lower).}
\label{fig:UnitCells}
\end{figure}

We will now consider how accurate the permittivity and permeabilities from current-driven homogenization (Sec. \ref{sec:four_mus}) predict the reflection of the semi-infinite periodic structure (Sec. \ref{sec:semiinfinit}). The reflection coefficient \eqref{rTM} is calculated using the permittivity \eqref{epsCDH} paired with the four different permeabilities ($\mull$, $\mumm$, $\muvy$, and $\mutl$). The results are compared with FDTD simulations of the same semi-infinite structure. We used the free open source software Meep for FDTD simulations. We do this for the three materials with unit cells shown in Fig.~\ref{fig:UnitCells}.

According to an earlier investigation of the current driven homogenization method for a 1D layered structure \cite{Markel13}, the permeability $\mutl$ does not lead to a more accurate reflection coefficient than that resulting from a trivial $\mull=1$. We verify this conclusion in Subsec.~\ref{sub:layers}. However, in Subsecs.~\ref{sub:twobars} and \ref{sub:splitring} we demonstrate that this is not the case for the 2D metal two-bar and split-ring structures, which more closely mimic the magnetism of natural media.

In the FDFD calculations a quadratic grid of $600\times 600$ points was used to represent the unit cells of all three examples. Grids of $200\times 1$ and $150\times\ 150$ points were used to represent the unit cells in the FDTD simulations for the 1D and 2D examples, respectively. In all simulations we have used a unit cell size $a=1~\mu$m. We note that the resonance of the split-ring structure is very sensitive to the resolution and the numerical representation of the gap. A large resolution was therefore necessary to obtain a convergent result in Figs.~\ref{fig:splitring_refl}-\ref{fig:splitring_mus}.

\subsection{Layered structure}\label{sub:layers}
\begin{figure}[!t]
\center
\includegraphics[width=0.4\textwidth]{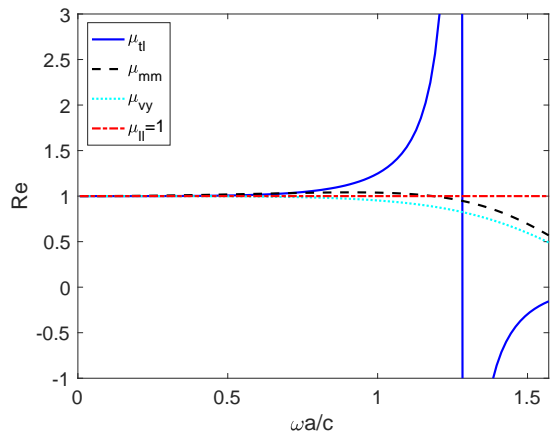}
\caption{The four $\mu$'s of the layered structure. We plot only the real parts, as the imaginary parts are negligible for all four $\mu$'s.} 
\label{fig:layers_mus}
\end{figure}

The unit cell for the layered structure is shown in the upper part of Fig.~\ref{fig:UnitCells}. The layers are dielectric with $\varepsilon=16$.
\begin{figure}[!t]
\center
\includegraphics[width=0.4\textwidth]{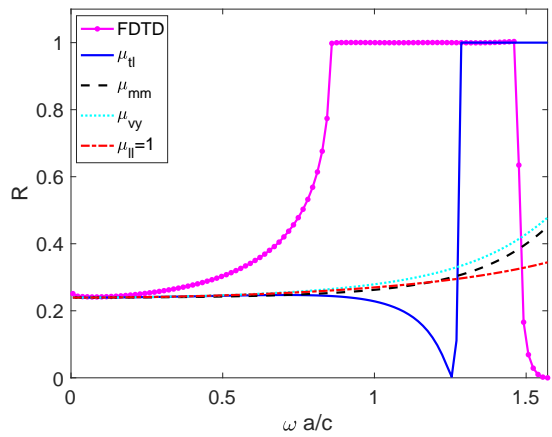}
\caption{Power reflection $R=|\rTM|^2$ for the layered dielectric structure obtained from FDTD, and from \eqref{rTM} using the different permeabilities.} 
\label{fig:layers_refl}
\end{figure}
The power reflection coefficients $R=|\rTM|^2$ obtained from FDTD and the Fresnel equation \eqref{rTM} using the four different permeabilities are shown in Fig. \ref{fig:layers_refl}. The figure shows that the Fresnel and FDTD reflection curves match well only for very small frequencies ($\omega a/c \lesssim 0.2$). As shown in Fig. \ref{fig:layers_mus} the four permeabilities all agree on an approximately non-magnetic response ($\mu=1$) in this frequency region. This is in line with the previous analysis \cite{Markel13} of a similar structure.

Local effective parameters are justified in Appendix~\ref{app:locality} to approximate the macroscopic behavior of the infinite layered structure for $\omega a/c \lesssim 0.6$. None of the four Fresnel reflection coefficients do however predict the actual (FDTD) reflection well in the region $0.3 < \omega a/c < 0.6$. This suggests that the conventional Maxwell boundary conditions do not apply, for the macroscopic fields defined as the fundamental Floquet modes. This peculiarity is further discussed in Sec.~\ref{sec:discussion}.

\subsection{Silver two-bars}\label{sub:twobars}
The unit cell for the silver two-bar structure is shown in the middle of Fig.~\ref{fig:UnitCells}. The silver in this and the next example is described by a Drude--Lorentz model for Ag with parameters from Ref. \cite{rakic98}

\begin{figure}[t]
\center
\includegraphics[width=0.4\textwidth]{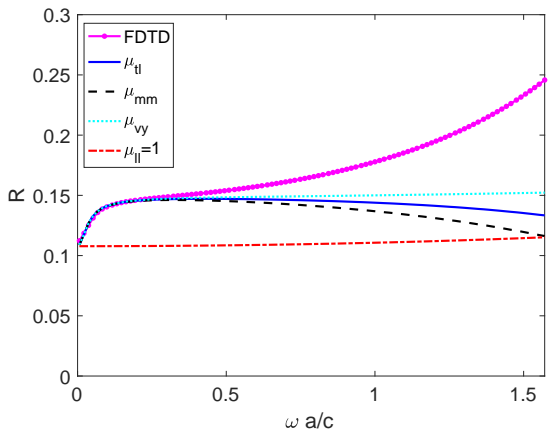}
\caption{Power reflection for the silver two-bar medium obtained from FDTD, and from \eqref{rTM} using the different $\mu$'s.}
\label{fig:twobars_refl}
\end{figure}
\begin{figure}[t]
	\center
	\subfloat{
		\includegraphics[width=0.4\textwidth]{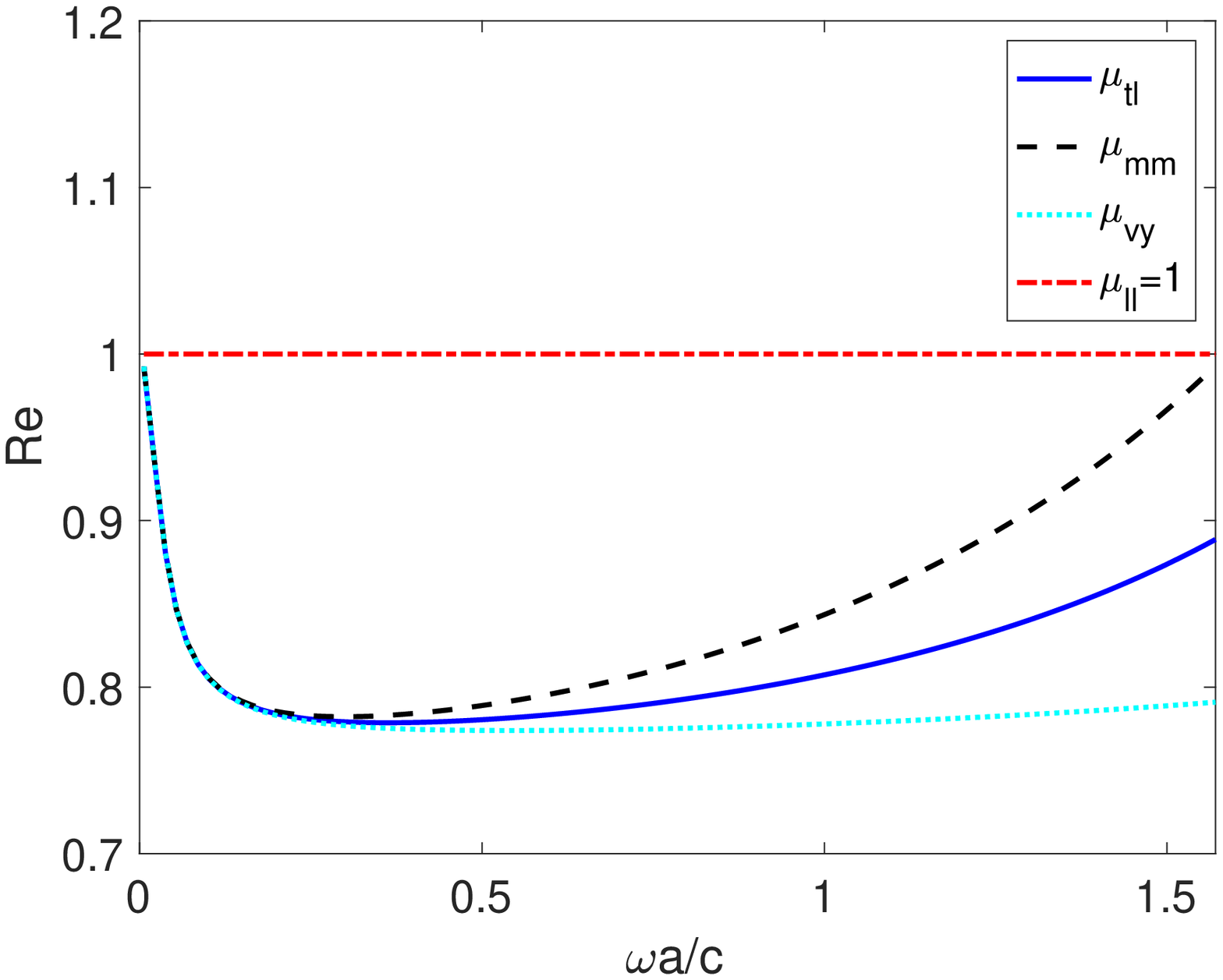}} \\
	\subfloat{
		\includegraphics[width=0.4\textwidth]{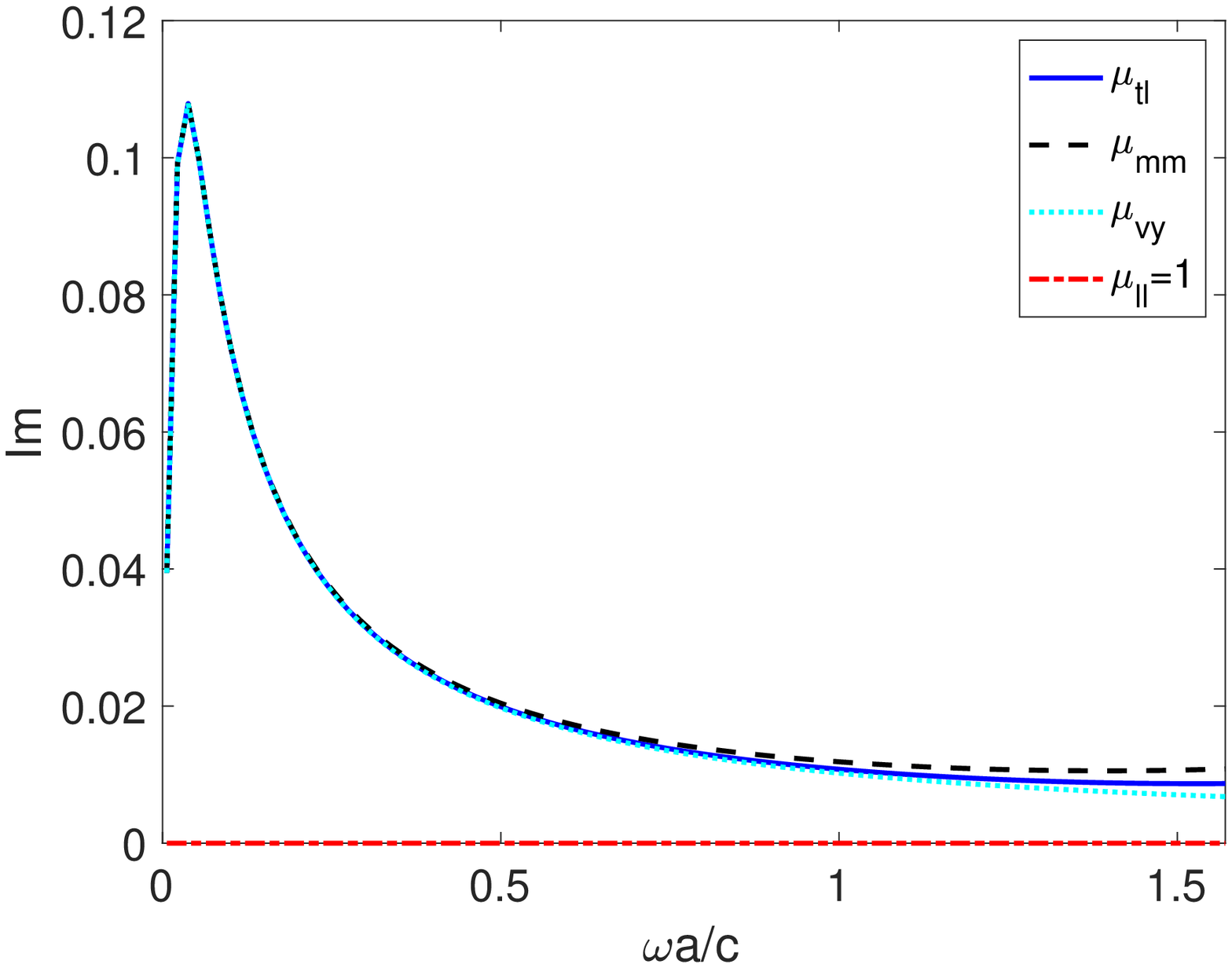}}
	\caption{The four $\mu$'s of the two-bar metamaterial; real parts (upper plot) and imaginary parts (lower plot). The bars are made from silver, and $a = 1~\mu$m.}
	\label{fig:twobars_mus}
\end{figure}

It is seen from Fig.~\ref{fig:twobars_refl} that \eqref{rTM} with the permeabilities $\mutl$, $\mumm$ and $\muvy$ predicts the actual reflection well for the frequency range $\omega a/c \lesssim 0.4$. As shown in Fig.~\ref{fig:twobars_mus} these three permeabilities coincide, and describe the medium as diamagnetic in this frequency region. The trivial $\mu_\text{ll}=1$ does however give poor results in the region. For the silver two-bar medium, the permeability obtained from current driven homogenization thus contribute to predict the reflection accurately, in a frequency region with a significant magnetic response.

\subsection{Silver split-ring}\label{sub:splitring}
\begin{figure}[h]
\center
\includegraphics[width=0.4\textwidth]{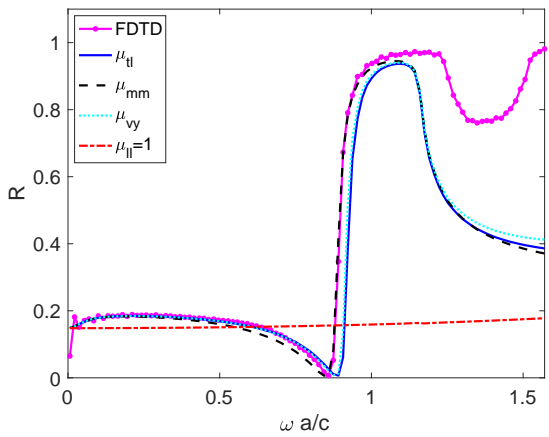}
\caption{Power reflection for the silver split-ring medium obtained from FDTD, and from \eqref{rTM} using the four different $\mu$'s.}
\label{fig:splitring_refl}
\end{figure}

\begin{figure}[t]
\center
 \subfloat{
\includegraphics[width=0.4\textwidth]{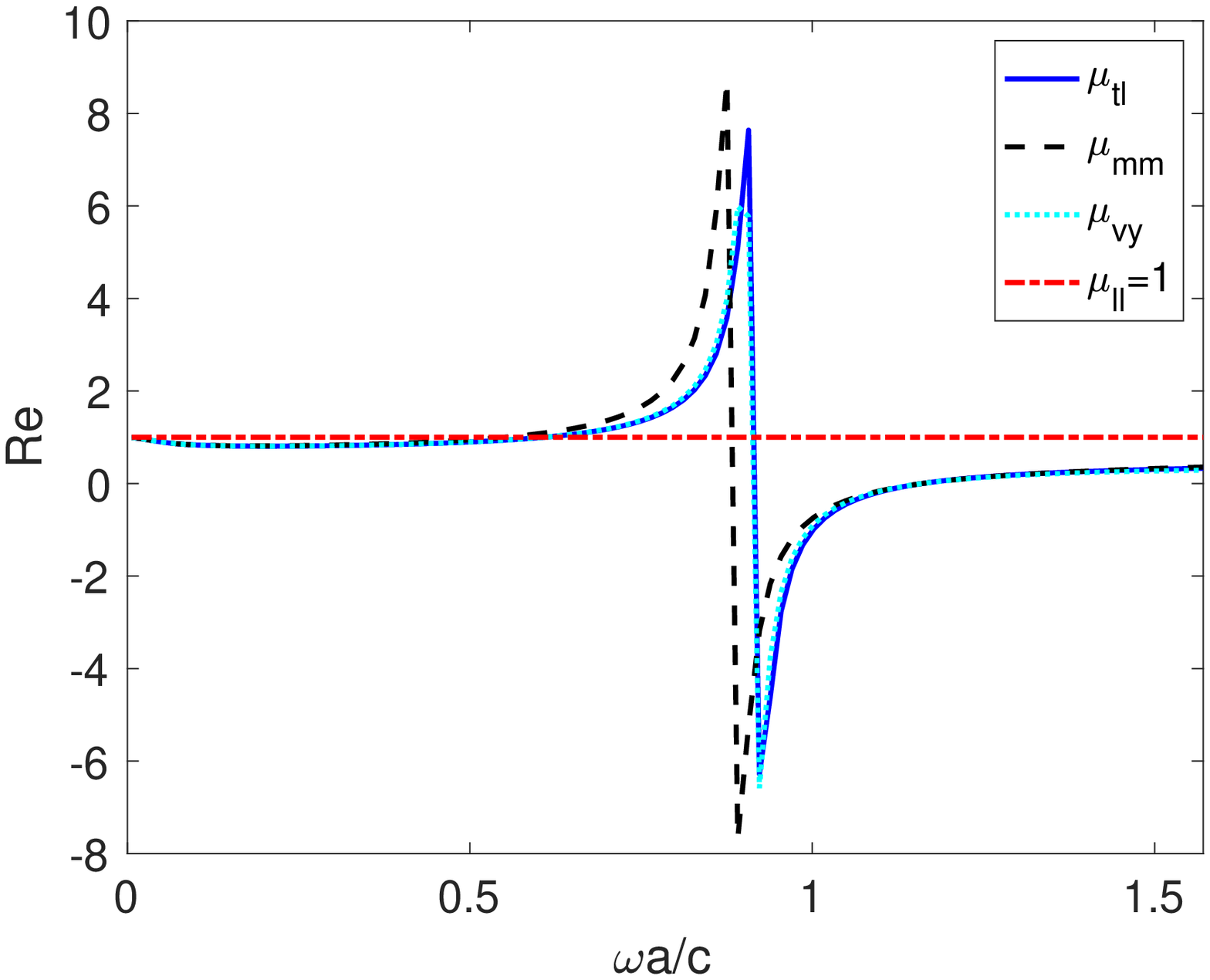}} \\
 \subfloat{
\includegraphics[width=0.4\textwidth]{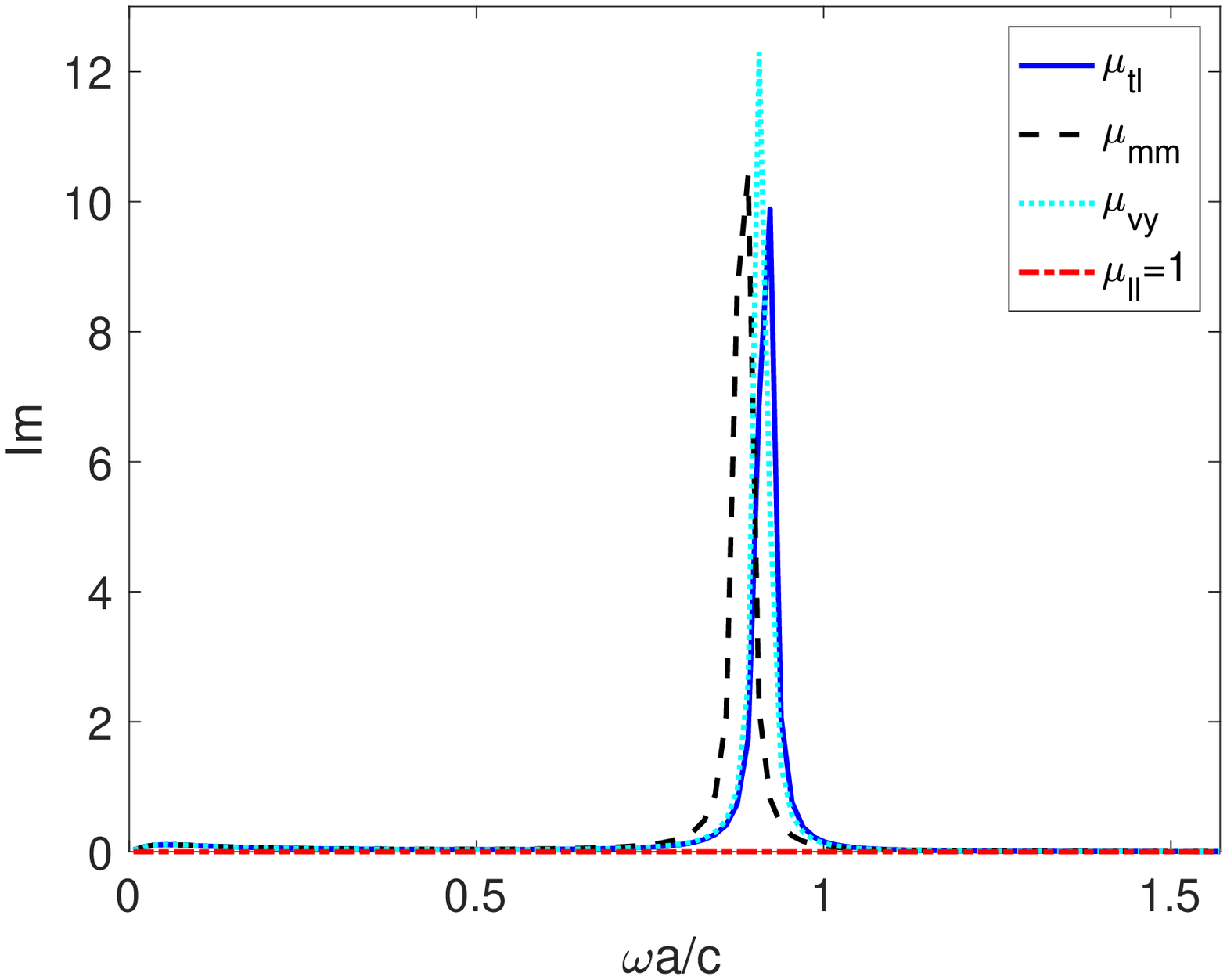}}
\caption{The four $\mu$'s of the silver split-ring metamaterial, $a = 1~\mu$m. Real parts (upper plot) and imaginary parts (lower plot).}
\label{fig:splitring_mus}
\end{figure}

The unit cell for the silver split-ring structure is shown in the lower part of Fig.~\ref{fig:UnitCells}. It is seen from Fig.~\ref{fig:splitring_refl} that the Fresnel coefficient with the permeabilities $\mu_\text{tl} \approx \mu_\text{mm} \approx \mu_\text{vy}$ predicts the FDTD coefficient quite well for frequencies $\omega a/c \lesssim 1.1$. As seen in Fig.~\ref{fig:splitring_mus} all four permeabilities are close to $1$ for $\omega a/c \lesssim 0.5$. However, even in this frequency range the diamagnetic response causes the actual reflection to deviate from the prediction by the trivial $\mu_\text{ll} = 1$. For $0.6 < \omega a/c < 1$ the three nontrivial permeabilities suggest a strong magnetic response (resonance), and the prediction based on the trivial $\mu_\text{ll}$ thus becomes very poor.

\section{Discussion and conclusion}\label{sec:discussion}
The standard derivation of the Fresnel reflection coefficient \eqref{rTM} is based on the continuity of the tangential components of the macroscopic $\mc{E}$ and $\mc{H}$ fields, together with the assumption of a local medium. For sufficiently high frequencies the locality condition \eqref{epsOK} is not fulfilled, and there is no reason to expect that \eqref{rTM} will predict the reflection from the periodic structure.

Nevertheless, for the 2D examples in Subsecs. \ref{sec:examples}~\ref{sub:twobars}-\ref{sub:splitring}, the Fresnel reflection coefficient \eqref{rTM}, calculated with the permittivty \eqref{epsCDH} and any of the nontrivial permeabilities $\mumm$, $\muvy$, or $\mutl$, matches the FDTD result fairly well in a frequency range with a nontrivial magnetic response.

The examples in Sec.~\ref{sec:examples} show that the Fresnel coefficient predicts the reflection well only in frequency ranges where the three nontrivial permeabilities coincide. This condition means, in somewhat loose terms, that the magnetic moment density dominates the $\mathcal{O}(k^2)$-term of $\vek\epsilon(\omega,\vek k)$.

We have also seen that local constitutive parameters $\epseff$ and $\mueff$ describing the Landau--Lifshitz permittivity $\epsOK$ well is not a sufficient condition for the standard Fresnel equation \eqref{rTM} to be valid. This is particularly clear from the 1D example where the Fresnel reflectivity did not match the exact result, except for very small frequencies where the magnetic response is absent. This mismatch happened even in a frequency range where the medium was verified to be local.

Thus, although a medium is local, \eqref{rTM} is sometimes not able to predict the actual reflection from the metamaterial. This means that continuity of the tangential component of $\mc{E}$ and $\mc{H}$ at $z=0$ must be violated. Alternatively, there is a ``transition layer'' close to the boundary, where this discontinuity is smoothed out.

In the 1D case, we can define the macroscopic fields as a superposition of the fundamental Floquet modes, for $z>0$ and $z<0$ separately. Then, for $z>0$, $\mc{E}$ is given by \eqref{Emac}, while for $z<0$ we simply have $\mc{E} = \vek{e}$. Since the tangential component of $\vek{e}$ is always continuous we will only get continuity of the tangential component of $\mc{E}$ if $\mc{E}(z=0^+) = \vek{e}(z=0)$, which there is no reason should be the case in general. The discontinuity is seen explicitly by \eqref{disE} and \eqref{disB} in Appendix \ref{app:newRefl}. For normalized, incident electric field, the discontinuity of the macroscopic electric field is given by the macroscopic transmission coefficient times $1- u_e(0)/\bar u_e$. Here $u_e(z)$ is the Bloch function, and $\bar u_e$ its average over a unit cell.

We could have defined macroscopic fields differently, e.g. using test function averaging \cite{russakoff,jackson,silveirinha09}. This leads to continuous fields; however at the expense of introducing a transition layer where the constitutive parameters $\epsilon$ and $\mu$ are not valid. Of course, such a definition will not make the Fresnel equation more accurate. The microscopic fields are the same, and the amount of each approximate-plane-waves far away from the boundary, will not change.

The Maxwell boundary conditions for the macroscopic fields can be restored by introducing macroscopic surface currents: 
\begin{subequations}
\begin{align}\label{EBtcond}
    & \left[\mc{E}(z=0^-) - \mc{E}(z=0^+)\right]\times\vekh z = -\mc K_\text{m}, \\
    & \left[\mc{B}(z=0^-) - \mc{B}(z=0^+)\right]\times\vekh z = \mc K_\text{e}.
\end{align}
\end{subequations}
Here $\mc K_\text{e}$ and $\mc K_\text{m}$ are macroscopic electric and magnetic surface currents, respectively. These surface currents are somewhat artificial, since they do not correspond to any physical, microscopic currents close to the boundary.

\appendix

\section{Locality and macroscopic wavenumber}\label{app:locality}
Under source-free propagation inside an infinite periodic structure the microscopic electric field of a mode can be written in Bloch form as in \eqref{emic}, but where the Bloch wave vector must satisfy a dispersion relation $\vek k= \kB(\omega)$. Here $\omega$ is the angular frequency of the incoming wave, and the dispersion relation $\vek k= \kB(\omega)$ may be found by insertion of the microscopic field \eqref{emic} into the source-free microscopic Maxwell's equations. In the case of source-free propagation, the spatial variation of the macroscopic fields \eqref{Emac} is thus given by the Bloch wave vector $\kB(\omega)$ rather than the forced $\vek{k}$ of the plane wave source in the current driven homogenization method.

If the microscopic susceptibility $\varepsilon(\vek{r})-1$ and/or the frequency $\omega$ is large, the assumption of small $ka$ is not necessarily valid. Then the Landau--Lifshitz permittivity $\vek\epsilon(\omega,\vek k)$ will contain terms $\bigO(k^4)$, and \eqref{epsOK} is not satisfied. To verify whether the locality condition is fulfilled for the media and frequency ranges in Sec. \ref{sec:examples}, we plot the left and right hand sides of the $xx$ tensor element of \eqref{epsOK} as a function of $\omega$. In these plots we use all four permeabilities, and set $\vek{k} = \kB(\omega)$. As expected the right hand side of \eqref{epsOK} calculated using $\mutl$ gives best match with the left hand side, as this is the permeability which includes as much as possible of the $\mathcal{O}(k^2)$-term of $\vek\epsilon(\omega,\vek k)$.

We also check for which frequency ranges $k_\text{B}$ can be approximated by $k=\sqrt{\epsilon\mu}\omega/c$. Here the permittivity \eqref{epsCDH} is paired with the four permeabilities, respectively, giving four different wavenumbers. 

\begin{figure}[t]
\center
\includegraphics[width=0.4\textwidth]{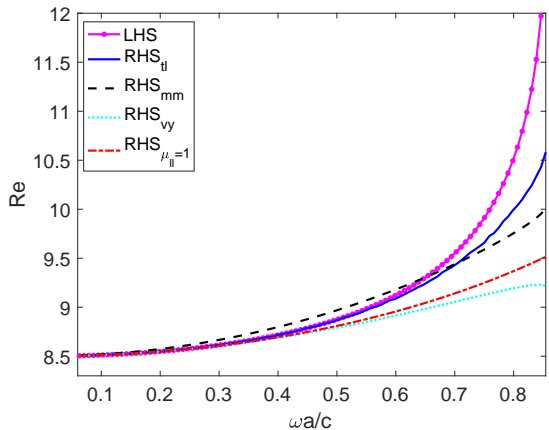}
\caption{The left hand side versus the right hand side of \eqref{epsOK} for $\vek{k} = \kB(\omega)$ for the layered structure.  We plot only the real parts, as the imaginary parts are negligible.}
\label{fig:locality_layers}
\end{figure}
\begin{figure}[t]
\center
\includegraphics[width=0.4\textwidth]{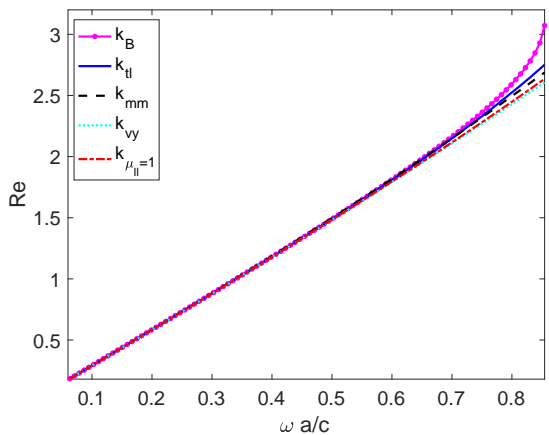}
\caption{Comparison of the wavenumbers for the layered structure. We plot only the real parts, as the imaginary parts are negligible.}
\label{fig:kz_layers}
\end{figure}

As seen in Fig.~\ref{fig:locality_layers} the locality condition \eqref{epsOK} with the parameters $\epsilon$ and $\mutl$ is a good approximation for $\omega a/c \lesssim 0.6$. For larger frequencies the left and right hand sides of \eqref{epsOK} starts to deviate, which means the approximation of weak spatial dispersion does not hold for such high frequencies. Since this is a 1D structure, only a single Bloch mode will be excited. 

Note that exact values for the Bloch wavenumbers $k_\text{B}$ may be obtained from \eqref{Blochk}. Since the value $k_\text{B}$ is found by inverse cosine of \eqref{Blochk} the branch for $k_\text{B}$ must be chosen. We choose $k_\text{B}$ such that $\Im\, k_\text{B} > 0$ and $\Re\, k_\text{B} < \pi/a$ for all $\omega > 0$. No matter which branch is chosen,
\eqref{epsOK} does however not give an accurate description for $\omega a/c > 0.6$. From Fig.~\ref{fig:layers_refl} it is seen that this is where the first Bragg resonance occurs. For such high frequencies it is not possible to describe the periodic structure in terms of effective parameters $\epseff$ and $\mueff$ \cite{Simovski09b}.

From Fig.~\ref{fig:kz_layers} it is seen that $\sqrt{\epseff\mueff}\omega/c$ using any of the four $\mu$'s approximates the Bloch wavenumber $k_\text{B}$ well for $\omega a/c \lesssim 0.7$. This means that the dispersion relation may be  approximated by $k^2 = \epseff\mueff \omega^2/c^2$ for these frequencies. 

\section{Mode matching}\label{app:newRefl}
Consider a semi-infinite medium with 2D periodicity. The boundary of the semi-infinite medium is $z=0$, and the medium is homogeneous along the $y$-direction. A plane-wave source with magnetic field amplitude 1 and transversal dependence $\exp(ik_xx)$ is located somewhere for $z<0$. The magnetic field points in the $\vekh y$-direction (TM polarization). If the periodic medium were infinite, the solutions could have been written in Bloch form
\be\label{blochxz}
b^l(x,z) = u_b^l(x,z)\e{ik_x x+ik_z^l z},
\ee
with Bloch function $u_b^l(x,z)$. Here $k_x$ and $k_z^l$ satisfy a dispersion relation $\omega=\omega(k_x,k_z^l)$, which can be found by substituting \eqref{blochxz} into the source-free Maxwell equations. The superscript $l$ indexes the branches of the dispersion relation for the given $\omega$ and $k_x$. Although the medium now is semi-infinite, we still look for a solution which is a superposition of modes \eqref{blochxz}. As long as we are able to satisfy the boundary conditions at $z=0$, this strategy must be ok.

Let $K_k$ be the reciprocal lattice vector component along the $x$-direction. The magnetic field for $z<0$ is given by
\be\label{hzl}
b(x,z)=\e{ik_xx+ik_z(k_x)z} + \sum_k \rho_k \e{i(k_x+K_k)x-ik_z(k_x+K_k)z},
\ee
where $\rho_k$ denote the reflection coefficients, and $k_z(k_x)=(\omega^2/c^2-k_x^2)^{1/2}$. For $z\geq 0$ we have
\be\label{hzg}
b(x,z) = \sum_{l} \tau_l u_b^l(x,z)\e{ik_xx+ik_z^lz},
\ee
where $\tau_l$ are the transmission coefficients.

The Maxwell boundary conditions require $b(x,z)$ to be continuous (recall that the microscopic inclusions are nonmagnetic): 
\be
\e{ik_xx} + \sum_k \rho_k \e{i(k_x+K_k)x} = \sum_{l} \tau_l u_b^l(x,0)\e{ik_xx}.
\ee
The periodic function $u_b^l$ can be written as a Fourier series along the $x$-direction:
\be\label{ubkl}
u_b^l(x,z)=\sum_k B_k^l(z) \e{iK_k x},
\ee
where $B_k^l(z)$ are the Fourier coefficients. This gives
\be
\e{ik_xx} + \sum_k \rho_k \e{i(k_x+K_k)x} = \sum_{l,k} \tau_l B_k^l(0)\e{i(k_x+K_k)x},
\ee
which after multiplication by $\e{-ik_xx-iK_nx}$, and integration wrt. $x$, can be written
\be
\delta_{n0} +  \rho_n = \sum_{l} \tau_l B_n^l(0).
\label{bboundc}
\ee

Similarly, one can express the continuity of the tangential electric field. The resulting system of two mode matching equations can be solved for the reflection and transmission coefficients.
The resulting, microscopic field is given by \eqref{hzl} and \eqref{hzg}. 

The longitudinal wavenumber $k_z(k_x+K_k)$ is imaginary for all $k$'s except $k=0$. Also, it is reasonable to assume that $k_z^l$ has a large imaginary part for all $l$'s except one, say $l=0$. Thus, sufficiently far away from the boundary we have
\be\label{bz0}
b(x,z)=
\begin{cases}
\e{ik_xx+ik_z(k_x)z} + \rho_0 \e{ik_xx-ik_z(k_x)z}, & z<0, \\
\tau_0 u_b^0(x,z)\e{ik_xx+ik_z^0z}, & z\geq 0.
\end{cases}
\ee

We now specialize to 1D, where the structure and source are homogeneous along the $x$-direction. This case is both TE and TM simultaneously. Writing the electric field in Bloch form $e(z)=u_e(z)\e{ik_\text{B}z}$ for $z>0$, the boundary conditions for the microscopic fields become
\begin{subequations} \label{boundc1d}
\begin{align}
& 1 + \rho_e = \tau_e u_e(0), \\
& ik_0 \left(1 - \rho_e\right) = \tau_e \left[ ik_\text{B} u_e(0) + u_e'(0) \right],
\end{align}
\end{subequations}
where $k_0$ is the wavenumber in vacuum, $k_\text{B}$ is the Bloch wavenumber in the structure, and $\rho_e$ and $\tau_e u_e(0)$ are the reflection and transmission coefficients for the microscopic electric field. The solutions for $\tau_e$ and $\rho_e$ become
\begin{subequations} \label{fresn12m_appe}
\begin{align}
& \tau_e u_e(0) =\frac{2k_0}{k_0+k_\text{B}-i\frac{u_e'(0)}{u_e(0)}}, \\
& \rho_e = \tau_e u_e(0)- 1 = \frac{k_0-k_\text{B}+i\frac{u_e'(0)}{u_e(0)}}{k_0+k_\text{B}-i\frac{u_e'(0)}{u_e(0)}} \label{newRefle}.
\end{align}
\end{subequations}
Similarly, we can find the reflection and transmission coefficients for the magnetic $b$ field (in an obvious notation): 
\begin{subequations} \label{fresn12m_appb}
\begin{align}
& \tau_b u_b(0) =\frac{2k_0\varepsilon(0^+)}{k_0\varepsilon(0^+)+k_\text{B}-i\frac{u_b'(0)}{u_b(0)}}, \\
& \rho_b = \tau_b u_b(0)- 1 = \frac{k_0\varepsilon(0^+)-k_\text{B}+i\frac{u_b'(0)}{u_b(0)}}{k_0\varepsilon(0^+)+k_\text{B}-i\frac{u_b'(0)}{u_b(0)}} \label{newReflb}.
\end{align}
\end{subequations}
Here $\varepsilon(0^+)$ is the microscopic permittivity immediately to the right of the boundary ($z=0^+)$. Note that we must have $\rho_b=-\rho_e$.

In light of \eqref{Emac} we identify the macroscopic fields 
\be
    \mathcal E(z) =
    \begin{cases}
        e(z), & \text{ for } z \leq 0, \\
        \tau_e\bar u_e\e{ik_\text{B}z}, & \text{ for } z>0,
    \end{cases}
\ee
\be
    \mathcal B(z) =
    \begin{cases}
        b(z), & \text{ for } z \leq 0, \\
        \tau_b \bar u_b\e{ik_\text{B}z}, & \text{ for } z>0,
    \end{cases}
\ee
where the bar denotes averaging over a unit cell. This means that $\rho_e$ and $\rho_b$ will remain reflection coefficients for the macroscopic fields, while the transmission coefficients will be different. Surprisingly, this leads to a discontinuity for the macroscopic electric field:
\be\label{disE}
    \mathcal E(0^+) - \mathcal E(0^-) = 
    \tau_e \bar u_e - (1+\rho_e) = \tau_e \Big[\bar u_e - u_e(0) \Big]. 
\ee
Similarly, we find
\be\label{disB}
    \mathcal B(0^+) - \mathcal B(0^-) = 
    \tau_b \bar u_b - (1+\rho_b) = \tau_b \Big[\bar u_b - u_b(0) \Big]. 
\ee

\def\cprime{$'$}

\end{document}